\documentclass{article}
\expandafter\let\csname opt@amsmath.sty\endcsname\relax
\makeatother
\usepackage{graphicx}
\usepackage{subfig}
\usepackage{url}

\usepackage{breakurl}
\usepackage{amsmath}
\usepackage[numbers]{natbib}
\usepackage[english]{babel}
\usepackage[ruled,vlined]{algorithm2e}
\usepackage{multirow}
\usepackage{flushend}
\usepackage[a4paper, total={6in, 8in}]{geometry}
\begin{document}
\title{\textbf{A Deep Learning Based DDoS Detection System in Software-Defined Networking (SDN)}}
\author{ \small{Quamar Niyaz}\thanks{Corresponding author: \texttt{quamar.niyaz@utoledo.edu}}, \small{Weiqing Sun}, \small{Ahmad Y Javaid}\\ \small{\texttt{\{quamar.niyaz, weiqing.sun, ahmad.javaid\}@utoledo.edu}}\\\small{College of Engineering}\\ \small{The University of Toledo}\\ \small{Toledo, OH-43606, USA}}
\date{}
\maketitle
\begin{abstract}
\noindent
Distributed Denial of Service (DDoS) is one of the most prevalent attacks that an organizational network infrastructure comes across nowadays. We propose a deep learning based multi-vector DDoS detection system in a software-defined network (SDN) environment. SDN provides flexibility to program network devices for different objectives and eliminates the need for third-party vendor-specific hardware. We implement our system as a network application on top of an SDN controller. We use deep learning for feature reduction of a large set of features derived from network traffic headers. We evaluate our system based on different performance metrics by applying it on traffic traces collected from different scenarios. We observe high accuracy with a low false-positive for attack detection in our proposed system. 

\end{abstract}
\section{Introduction}
\label{intro}
Distributed denial of service (DDoS) attacks results in unavailability of network services by continuously flooding its servers with undesirable traffic. Low-price Internet subscriptions and readily available attack tools led to a vast increase in volume, size, and complexity of these attacks in the recent past. According to the forecast of Cisco Visual Networking Index (VNI)~\cite{cisco}, DDoS incidents will reach up to 17 million in 2020, a threefold increment compared to 2015. The nature of attacks has also changed to being multi-vector rather than having a single type of flooding. A study reported that 64\% attacks until mid-2016 were multi-vectors that include TCP SYN floods and DNS/NTP amplification combined together~\cite{verisign}. Adversaries or hacktivists use DDoS attacks for extortion, revenge, misguided marketing, and online protest. Many financial, public sector, media, and social entertainment sites are recent victims~\cite{neustar,bbc,blizzard} and suffered from monetary and reputation damages. Therefore, detection and mitigation of these attacks in real-time have become a prime concern for large organizations.

Recently, both software-defined networking (SDN) and deep learning (DL) have found several useful and interesting applications in the industry as well as the research community. SDN provides centralized management, global view of the entire network, and programmable control plane; makes network devices flexible for different applications. These features of SDN offer better network monitoring and enhanced security of the managed network compared to traditional networks~\cite{scheh,shin}. On the other hand, DL based approaches outperformed existing machine learning techniques when applied to various classification problems. They improve feature extraction/reduction from a high-dimensional dataset in an unsupervised manner by inheriting the non-linearity of neural networks~\cite{schm}. Researchers have also started to apply DL for the implementation of various intrusion detection systems and observed desirable results discussed in Section~\ref{related}. In this work, we implement a DDoS detection system that incorporates stacked autoencoder (SAE) based DL approach in an SDN environment and evaluate its performance on a dataset that consists of normal Internet traffic and various DDoS attacks.

The organization of this paper is as follows. Section~\ref{related} discusses related work on DDoS detection in an SDN environment and use of DL for network intrusion detection. Section~\ref{background} gives an overview of SDN and SAE. In Section~\ref{system}, we discuss the architecture of our proposed system. Section~\ref{result} presents experimental set-up and performance evaluation of the system. Finally, Section~\ref{conclusion} concludes the paper with future work directions.

\section{Related Work}
\label{related}
We discuss the related work from two perspectives, one in which DL has been used for network intrusion detection and the other in which DDoS detection is addressed in an SDN environment.
\subsection{Intrusion detection using DL}
In~\cite{salama}, Mostafa et al. used deep belief network (DBN) based on restricted Boltzmann machine (RBM) for feature reduction with support vector machine (SVM) as a classifier to implement a network intrusion detection system (NIDS) on NSL-KDD~\cite{tavalla} intrusion dataset. In~\cite{fiore}, Ugo et al. used discriminative RBM (DRBM) to develop a semi-supervised learning based network anomaly detection and evaluated its performance in an environment where network traffic for training and test scenarios were different. They used real-world traffic traces and KDD Cup-99~\cite{kdd} intrusion dataset in their implementation. In~\cite{gao}, Gao et al. used RBM based DBN with a neural network as a classifier to implement an NIDS on KDD-Cup 99 dataset. In~\cite{kang}, Kang et al. proposed an NIDS for the security of in-vehicular networks using DBN and improved detection accuracy compared to previous approaches. In~\cite{javaid}, we implemented a deep learning based NIDS using NSL-KDD dataset. We employed self-taught learning~\cite{raina} that uses sparse autoencoder instead of RBM for feature reduction and evaluated our model separately on training and test datasets. In~\cite{ma}, Ma et al. proposed a system that combines spectral clustering (SC) and sparse autoencoder based deep neural network (DNN). They used KDD-Cup99, NSL-KDD, and a sensor network dataset to evaluate the performance of their model.

\subsection{DDoS detection in SDN environment}
In~\cite{braga}, Braga et al. proposed a light-weight DDoS detection system using self-organized map (SOM) in SDN. Their implementation uses features extracted from flow-table statistics collected at a certain interval to make the system light-weight. However, it has limitation in handling traffic that does not have any flow rules installed. In~\cite{giotis}, Giotis et al. combined an OpenFlow (OF) and sFlow for anomaly detection to reduce processing overhead in native OF statistics collection. As the implementation was based on flow sampling using sFlow, false-positive was quite high in attack detection. In~\cite{lim}, Lim et al. proposed a DDoS blocking application (DBA) using SDN to efficiently block legitimately looking DDoS attacks. The system works in collaboration with the targeted servers for attack detection. The prototype was demonstrated to detect HTTP flooding attack. In~\cite{mousavi}, Mousavi et al. proposed a system to detect DDoS attacks in the controller using entropy calculation. Their implementation depends on a threshold value for entropy to detect attacks which they select after performing several experiments. The approach may not be reliable since threshold value will vary in different scenarios. In~\cite{wang}, Wang et al. proposed an entropy based light-weight DDoS detection system by exporting the flow statistics process to switches. Although the approach reduces the overhead of flow statistics collection in the controller, it attempts to bring back the intelligence in network devices.

In contrast to the discussed work, we use SAE based DL model to detect multi-vector DDoS attacks in SDN. We use a set of large number of features extracted from network packet headers and then use DL to reduce this set in an unsupervised manner. We apply our model on traffic dataset collected in different environments. The proposed system attempts to detect attacks on both the SDN control plane and the data plane, and is implemented completely on the SDN controller.

\section{Background Overview}
\label{background}
We discuss SDN and SAE before describing our DDoS detection system.
\subsection{Software-Defined Networking (SDN)}
\label{sdn}
As discussed earlier, the SDN architecture decouples the control plane and data plane from network devices, also termed as `\textit{switches}', and makes them simple packet forwarding elements. The decoupling of control logic and its unification to a centralized controller offers several advantages compared to the current network architecture that integrates both the planes tightly. Administrators can implement policies from a single point, i.e. controller, and observe their effects on the entire network that makes management simple, less error-prone, and enhances security. Switches become generic and vendor-agnostic. Applications that run inside a controller can program these switches for different purposes such as layer 2/3 switch, firewall, IDS, load balancer using API offered by a controller to them~\cite{kreutz}.

\begin{figure*}[!ht]
\centering
\begin{tabular}{@{}ccc@{}}
\subfloat[Different planes and network applications in SDN]{\includegraphics[width=0.32\textwidth]{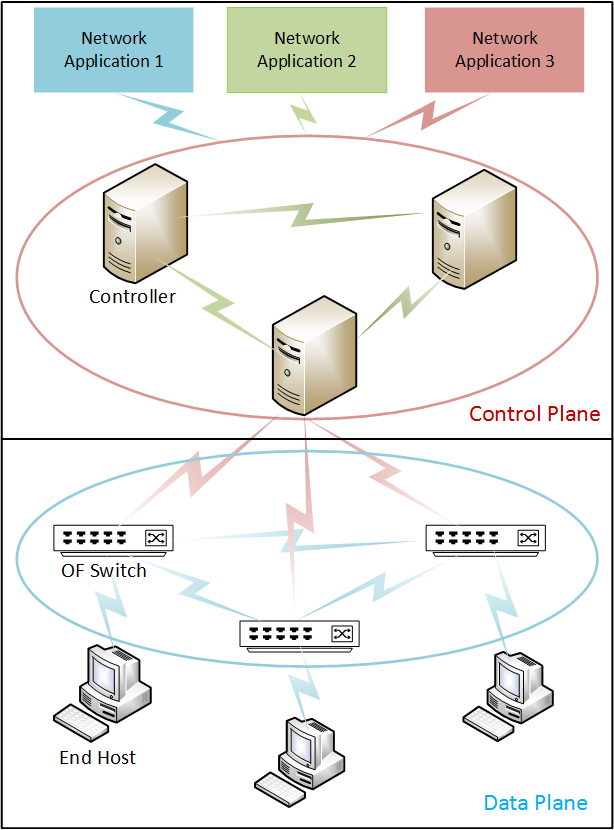}\centering\label{fig:sdn1}} & 
\subfloat[Reactive traffic flow set-up in SDN~\cite{niyaz}]{\includegraphics[width=0.45\textwidth]{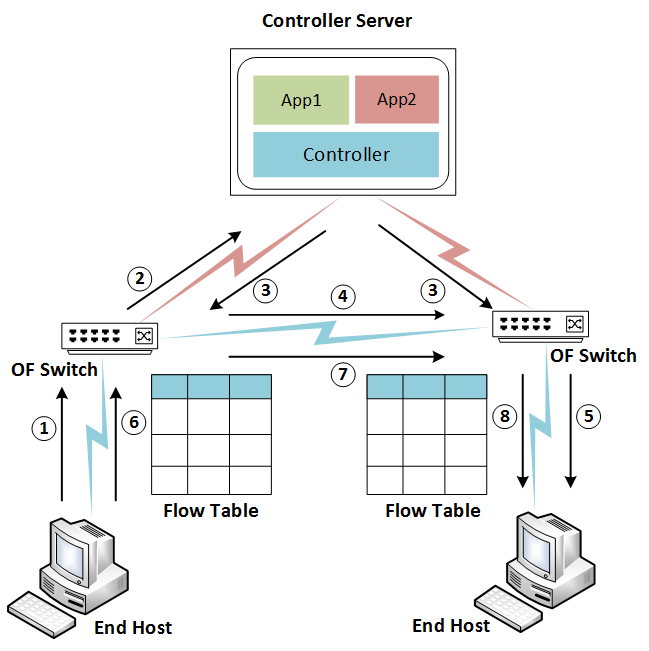}\label{fig:sdn2}} &
\end{tabular}
\caption[]{An SDN architecture and basic traffic flow in SDN}
\label{fig:sdn}
\end{figure*}

Figure~\ref{fig:sdn1} shows the SDN architecture with its different planes and applications. Switches, end hosts, and communication between them form the data plane. The controller is either a single server or a group of logically centralized distributed servers. The controller can run on commodity hardware and communicates with switches using standard APIs called southbound interfaces. One of the \textit{de facto} standards for southbound interfaces is OpenFlow (OF) protocol~\cite{mckeown}. The controller servers communicate with each other using east-westbound interfaces. Network applications communicate with the controller using northbound interfaces.

The controller and switches exchange various types of messages using OF protocol over either a TLS/SSL encrypted or open channel. These messages set-up switch connection with the controller, inquire network status or manage traffic flows in the network. Switches have flow tables for flow rules that contain match-fields, counters, and actions to handle traffic flows in the network. SDN defines \textit{flow} as a group of network packets that have same values for certain packet header fields. The controller installs flow rules for traffic flows based on the policies dictated by the network applications. 

Flow rule installation takes place in switches either in reactive mode or proactive mode. The reactive mode works as follows. When a packet enters a switch, it looks up for a flow rule inside its flow tables that matches with the packet headers. If a rule exists for the packet, the switch takes an action that may involve packet forwarding, drop or header modification. If a table-miss happens, i.e., there are no flow rules for an incoming flow, the switch sends a \textit{packet\_in} message to the controller that encapsulates packet headers for the incoming flow. The controller extracts packet headers from the received message and sends a \textit{packet\_out} or \textit{flow\_mod} message to switches for the received packet's flow. The controller installs flow rules inside switches using \textit{flow\_mod} messages and switches perform actions on subsequent packets of the installed flow, without forwarding them to the controller. Figure~\ref{fig:sdn2} demonstrates the reactive mode set-up in SDN. Flow rules may expire after a certain time to manage the limited memory size of switches. The controller does not install rules in switches, instead, it instructs them to forward the packet from a single or multiple port(s) using \textit{packet\_out} messages. In contrast, the controller pre-installs flow rules into switches in proactive mode.

\subsection{Stacked Autoencoder (SAE)}
\label{sae}
\begin{figure*}[!ht]
\centering
\includegraphics[scale=0.32]{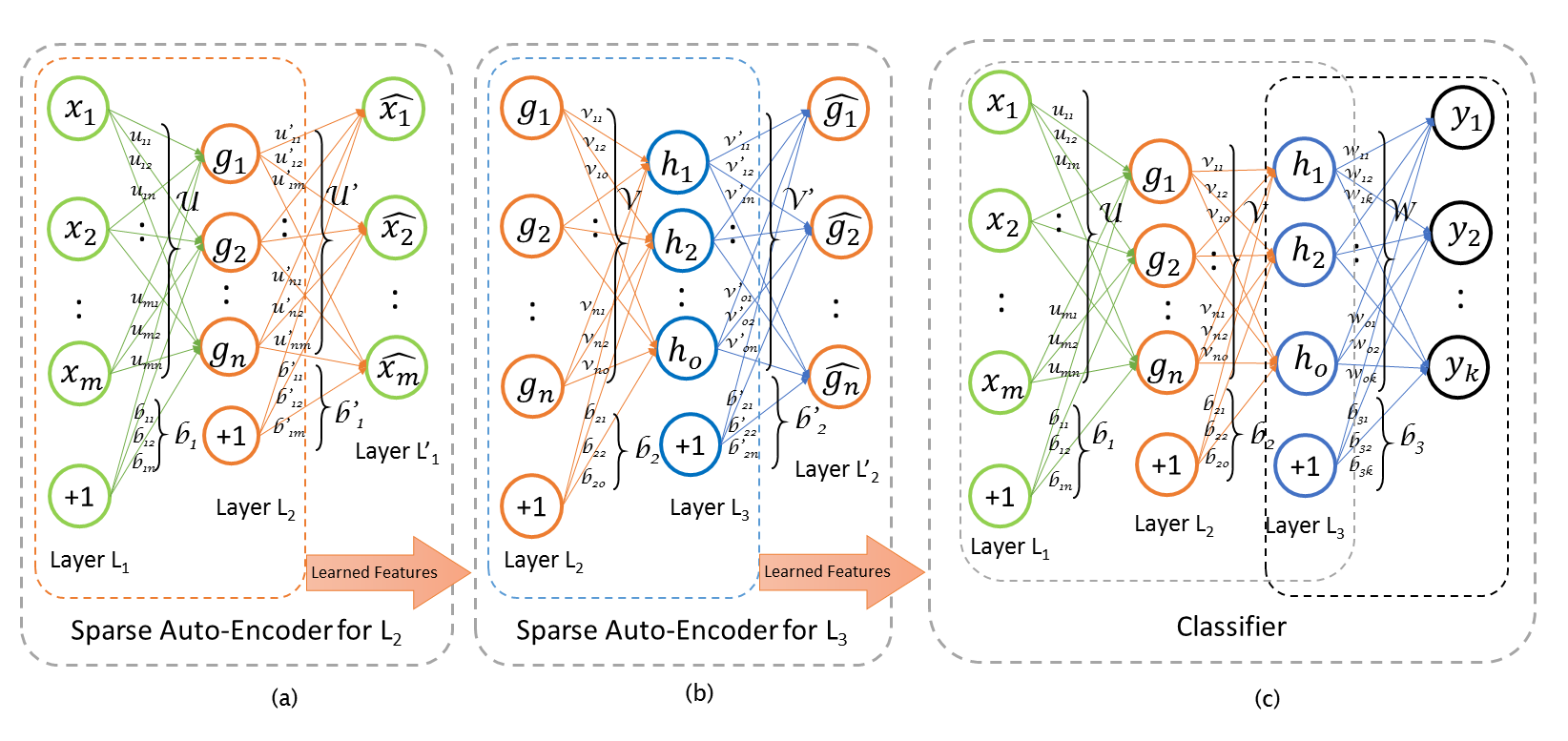}
\caption{A stacked autoencoder based deep learning model}
\label{fig:saefig}
\end{figure*}

Stacked Autoencoder (SAE) is a DL approach that consists of stacked sparse autoencoders and soft-max classifier for unsupervised feature learning and classification, respectively. We discuss sparse autoencoder before SAE. A sparse autoencoder is a neural network that consists of three layers in which the input and output layers contain $M$ nodes, and the hidden layer contains $N$ nodes. The $M$ nodes at the input represent a record with M features, i.e., $X=\{x_1, x_2, ..., x_m\}$. For the training purpose, the output layer is made an identity function of the input layer, i.e., $\Hat{X} = X$ shown in Figure~\ref{fig:saefig}a. The sparse autoencoder network finds optimal values of weight matrices, $U \in {\Re}^{N \times M}$ and $U' \in {\Re}^{M \times N}$, and bias vectors, ${b_1} \in {\Re}^{N \times 1}$ and ${b_1}' \in {\Re}^{M \times 1}$ while trying to learn an approximation of the identity function, i.e. ${\hat{X}} \approx X$ using back-propagation algorithm~\cite{ng2011sparse}. Many different functions are used for activation of hidden and output nodes, we use Sigmoid function, $g(z)=\frac{1}{1 + {e}^{-z}}$, for the activation of $g_{U,b_1}$ shown in Eqn.~\ref{eqn:sig}:
\begin{align}
    	g_{U,b_1}(X) = g(UX + b_1) = \frac{1}{1 + {e}^{-(UX +b_1)}}
    	\label{eqn:sig}
\end{align}
\setlength{\belowdisplayskip}{0pt} \setlength{\belowdisplayshortskip}{0pt}
\setlength{\abovedisplayskip}{0pt} \setlength{\abovedisplayshortskip}{0pt}
\begin{align*}
    J = \frac{1}{2r}\sum_{i=1}^{r}{\|X_i-\hat{X_i}\|^2} + \frac{\lambda}{2}(\sum_{n,m}{U}^2 +\sum_{m,n}{U'}^2
\end{align*} 

\begin{align}
   + \sum_{n}{b_1}^2 + \sum_{n}{b_1}'^2) + \beta\sum_{j=1}^{N} KL(\rho \| \hat{\rho}_j)
   \label{eqn:cost}
\end{align}  

Eqn.~\ref{eqn:cost} represents the cost function for optimal weight learning in sparse autoencoder. It is minimized using back-propagation. The first term in the RHS represents an average of sum-of-square errors for all the input values and their corresponding output values for all $r$ records in the dataset. The second term is a weight decay term with ${\lambda}$ as the decay parameter to avoid over-fitting. The last term is a sparsity penalty term that puts constraint on the hidden layer to maintain low average activation values and expressed using Kullback-Leibler (KL) divergence shown in Eqn.~\ref{eqn:KL}:
\begin{equation}
	\label{eqn:KL}
	KL(\rho \| \hat{\rho}_j) = \rho log\frac{\rho}{\hat{\rho}_{j}} + (1-\rho) log\frac{1-\rho}{1-\hat{\rho}_{j}} 
\end{equation}

where $\rho \in \{0,1\}$ is a sparsity constraint parameter and $\beta$ controls the sparsity penalty term. The $KL(\rho \| \hat{\rho}_j)$ becomes minimum when $\rho=\hat{\rho}_j$, where $\hat{\rho}_j$ is the average activation value of a hidden unit $j$ over all the training inputs.  

Multiple sparse autoencoders are stacked with each other in a way that the outputs of each layer is fed into the inputs of the next layer to create an SAE. Greedy-wise training is used to obtain optimal values of the weight matrices and bias vectors for each layer. For illustration, the first layer, ${g}$, on raw input $x$ is trained to obtain $U$, $U'$, ${b_1}$, ${b_1}'$. The layer $g$ encodes the raw input, $X$, using $U$ and $b_1$. Then, the encoded values are used as inputs to train the second layer to obtain parameters $V$, $V'$, $b_2$, ${b_2}'$ shown in Figure~\ref{fig:saefig}b. This process goes further until the last hidden layer is trained. The output of last hidden layer is fed into a classifier. Finally, all layers of SAE are treated as a single model and fine-tuned to improve the performance of the model shown in Figure~\ref{fig:saefig}c.
\section{Implementation of DDoS Detection System}
\label{system}
In an SDN, attacks can occur either on the data plane or control plane. Attacks on the former are similar to traditional attacks and affect a few hosts. However, attacks on the latter attempt to bring down the entire network. In this second kind of attack, adversaries fingerprint an SDN for flow installation rules and then send new traffic flows, resulting in flow table-misses in the switch~\cite{shin1}. This phenomenon forces the controller to handle every packet and install new flow rules in switches that consume system resources on the controller and switches. In our previous work ~\cite{niyaz}, we empirically evaluated the impact of SDN adversarial attacks on network services. In the current work, we implement a DDoS detection system as a network application in SDN to handle attacks for both cases.
\begin{figure}[!ht]
\centering
\includegraphics[scale=0.50]{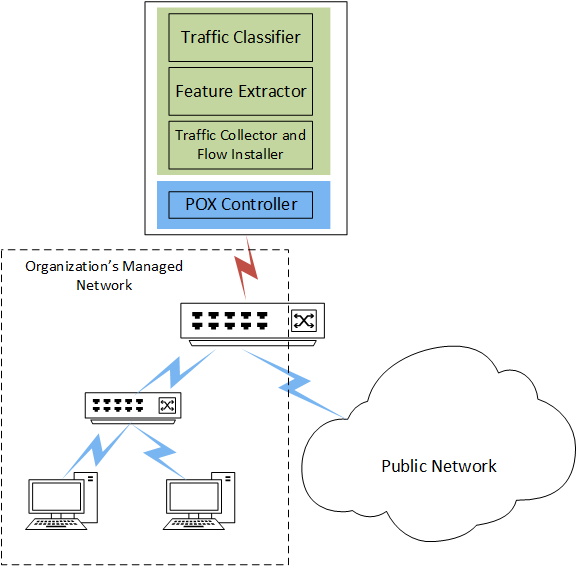}
\caption{A DDoS detection system implemented in SDN}
\label{fig:ids}
\end{figure}

\begin{table}
\centering
\begin{tabular}{ l|l|l|l }
  \hline
  \multicolumn{2}{c|}{\textbf{TCP}} & \textbf{UDP} & \textbf{ICMP} \\
  \hline
  \hline
  Src IP & Window & Src IP & Src IP \\
  Dst IP & SYN  & Dst IP & Dst IP\\
  Src Port & ACK & Src Port & ICMP Type \\
  Dst Port & URG & Dst Port & ICMP Code\\
  Protocol & FIN & Protocol  & Protocol\\
  Data Size & RST & Data Size & Data Size\\
  TTL & PUSH & TTL & TTL\\
  \hline
\end{tabular} 
\caption{Different headers extracted from TCP, UDP, and ICMP packets}           
\label{table:headers}
\end{table}
The detection system consists of three modules as shown in Figure~\ref{fig:ids}: i) Traffic Collector and Flow installer (\textbf{TCFI}), ii) Feature Extractor (\textbf{FE}), and iii) Traffic Classifier (\textbf{TC}). It should be emphasized here that to minimize false-positives, our system relies on every packet for flow computation and attack detection instead of sampling flows using some tools such as \emph{sFlow}.  

\begin{algorithm}
\DontPrintSemicolon
\KwData{Incoming network packets at the controller}
\KwResult{List of extracted packet headers for TCP, UDP, and ICMP}
\Begin{
    $packets\_list \longleftarrow \emptyset$\;
    $flows\_list \longleftarrow \emptyset$\;
    \While{$Timer$ for the FE is not triggered}{
        Receive a packet from switch\; 
        Store headers in $packets\_list$\;
        \If{Packet arrives due to flow table miss}{
            Compute $flow$ for the packet\;
            Compute symmetric flow, $symflow$, for $flow$\;
            \If{$symflow$ $\in$ $flows\_list$}{
                Remove $symflow$ from $flows\_list$\;
                Install flow rule for $symflow$ in switch(es)\;
                Install flow rule for $flow$ in switch(es)\;
            }\ElseIf{$flow$ $\notin$ $flows\_list$ }{
                Add $flow$ in $flows\_list$\;
                Output the packet to desired port\;
            }\Else{
                Output the packet to desired port\;
            }
        }
    }
}
\caption{TCFI Module\label{alg:tcfi}}
\end{algorithm}
\subsection{Traffic Collector and Flow Installer (TCFI)}
The TCFI module runs concurrently with the FE and TC modules which are triggered using a $timer$ function. It examines OF message type for an incoming packet at the controller. A message type determines the reason for a packet's arrival which is either due to a flow table-miss or an installed flow rule that forwards a packet towards the controller and desired physical ports. The TCFI extracts various header fields from a packet to identify its \textit{flow}. A \textit{flow} in TCP or UDP traffic is a group of packets having same values for protocol type, source and destination IP addresses, and source and destination port numbers. An ICMP \textit{flow} has similar header fields, except for port numbers it has ICMP message type and code. The TCFI extracts few more header fields from a packet that help in features extraction from flows. It stores all of these extracted headers in a list for every packet coming to the controller. Table~\ref{table:headers} shows the headers for TCP, UDP, and ICMP traffic that the TCFI extracts. It performs this task when a packet arrives due to pre-installed flow rules.

However, when a packet arrives due to a flow table-miss, it performs following tasks in addition to the one mentioned above. It looks up a symmetric flow corresponding to the packet's flow in the flow list. Two flows are \textit{symmetric} for TCP or UDP traffic if the source IP address and port number of one flow are similar to the destination IP and port number of the other, and vice-versa. For ICMP traffic, two flows are \textit{symmetric} if they are request and response types. If a symmetric flow exists for an incoming flow, then it installs forwarding rules for both of them in SDN switches and removes the symmetric one from the list. The rules include an action that forwards packets to desired physical ports and the controller for the incoming and its symmetric flows. The reason for installing rules only for symmetric flows is built on the assumption that attackers, in general, spoof their IP addresses to prevent responses towards them from victims. Therefore, the TCFI installs flow rules for legitimate traffic and avoids any flow table saturation attacks in switches. If it does not find any symmetric flow for an incoming packet, it looks up whether a flow already exists in the list for the same. If a flow exists, it forwards the packet from switches without installing any rules. Otherwise, it adds the packet's flow in the list and then forwards it. Algorithm~\ref{alg:tcfi} shows various steps involved in the TCFI. Although the algorithm appears similar to maximum entropy detector in ~\cite{mehdi}, i) it considers flow in general instead of flags based ii) a packet arrives at the controller either due to a table-miss or a forwarding rule towards the controller. iii) it stores packet headers for each packet arrives at the controller.  
\begin{table}
\centering
\begin{tabular}{ p{0.5 cm}|l }
  \hline
  \# & Feature Description\\
  \hline
  \hline
  1 & \# of incoming TCP flows\\
  2 & Fraction of TCP flows over total incoming flows \\
  3 & \# of outgoing TCP flows\\
  4 & Fraction of TCP flows over total outgoing flows \\
  5 & Fraction of symmetric incoming TCP flows \\
  6 & Fraction of asymmetric incoming TCP flows \\
  7 & \# of distinct src IP for incoming TCP flows\\
  8 & Entropy of src IP for incoming TCP flows \\
  9 & Bytes per incoming TCP flow \\
  10 & Bytes per outgoing TCP flow \\
  11 & \# of packets per incoming TCP flow \\
  12 & \# of packets per outgoing TCP flow \\
  13 & \# of distinct window size for incoming TCP flows \\
  14 & Entropy of window size for incoming TCP flows \\
  15 & \# of distinct TTL values for incoming TCP flows \\
  16 & Entropy of TTL values for incoming TCP flows \\
  17 & \# of distinct src ports for incoming TCP flows \\
  18 & Entropy of src port for incoming TCP flows\\
  19 & \# of distinct dst ports for incoming TCP flows \\
  20 & Entropy of dst ports for incoming TCP flows\\
  21 & Fraction of dst ports $\le$ 1024 for incoming TCP flows \\
  22 & Fraction of dst port $>$ 1024 for incoming TCP flows \\
  23 & Fraction of TCP incoming flows with SYN flag set \\
  24 & Fraction of TCP outgoing flows with SYN flag set \\
  25 & Fraction of TCP incoming flows with ACK flag set  \\
  26 & Fraction of TCP outgoing flows with ACK flag set \\
  27 & Fraction of TCP incoming flows with URG flag set \\
  28 & Fraction of TCP outgoing flows with URG flag set \\
  29 & Fraction of TCP incoming flows with FIN flag set \\
  30 & Fraction of TCP outgoing flows with FIN flag set \\
  31 & Fraction of TCP incoming flows with RST flag set \\
  32 & Fraction of TCP outgoing flows with RST flag set\\
  33 & Fraction of TCP incoming flows with PUSH flag set \\
  34 & Fraction of TCP outgoing flows with PUSH flag set \\
  \hline
\end{tabular} 
\caption{Features extracted for TCP flows}           
\label{table:tcpfeatures}
\end{table}
\begin{table}
\centering
\begin{tabular}{ p{0.5 cm}|l }
  \hline
  \# & Feature Description\\
  \hline
  \hline
  35 & \# of incoming UDP flows \\
  36 & Fraction of UDP flows over total incoming flows \\
  37 & \# of outgoing UDP flows\\
  38 & Fraction of UDP flows over total outgoing flows \\
  39 & Fraction of symmetric incoming UDP flows \\
  40 & Fraction of asymmetric incoming UDP flows \\
  41 & \# of distinct src IP for incoming UDP flows\\
  42 & Entropy of src IP for incoming UDP flows \\
  43 & Bytes per incoming UDP flow \\
  44 & Bytes per outgoing UDP flow \\
  45 & \# of packets per incoming UDP flow \\
  46 & \# of packets per outgoing UDP flow \\
  47 & \# of distinct src ports for incoming UDP flows \\
  48 & Entropy of src ports for incoming UDP flows \\
  49 & \# of distinct dst ports for incoming UDP flows \\
  50 & Entropy of dst ports for incoming UDP flows \\
  51 & Fraction of dst port $\le$ 1024  for incoming UDP flows \\
  52 & Fraction of dst port $>$ 1024  for incoming UDP flows \\
  53 & \# of distinct TTL values for incoming UDP flows\\
  54 & Entropy of TTL values for incoming UDP flows \\
  \hline
\end{tabular} 
\caption{Features extracted for UDP flows}           
\label{table:udpfeatures}
\end{table}
\begin{table}
\centering
\begin{tabular}{ p{0.5 cm}|l }
  \hline
  \# & Feature Description\\
  \hline
  \hline
  55 & \# of incoming ICMP flows\\
  56 & Fraction of ICMP flows over total incoming flows \\
  57 & \# of outgoing ICMP flows\\
  58 & Fraction of ICMP flows over total outgoing flows \\
  59 & Fraction of symmetric incoming ICMP flows \\
  60 & \# of asymmetric incoming ICMP flows \\ 
  61 & \# of distinct src IP for incoming ICMP flows \\
  62 & Entropy of src IP for incoming ICMP flows \\
  63 & Bytes per incoming ICMP flow \\
  64 & Bytes per outgoing ICMP flow \\
  65 & \# of packets per incoming ICMP flow \\
  66 & \# of packets per outgoing ICMP flow \\
  67 & \# of distinct TTL values for incoming ICMP flows\\
  68 & Entropy of TTL values for incoming ICMP flows \\
  \hline
\end{tabular} 
\caption{Features extracted for ICMP flows}           
\label{table:icmpfeatures}
\end{table}
\subsection{Feature Extractor (FE) and Traffic Classifier (TC)}
The detection system triggers the FE module using a $timer$ function. The FE takes packet headers from the packets list populated by the TCFI and extracts features from them for a set interval and resets the packet list to store headers for the next interval. Table~\ref{table:tcpfeatures},~\ref{table:udpfeatures}, and~\ref{table:icmpfeatures} show the list of 68 features that the FE extracts for TCP (34), UDP (20), and ICMP (14) flows, respectively. We derived this feature set after detailed literature survey and use SAE to reduce it. The FE computes these features for all hosts in a network which has incoming traffic flows for that particular interval. Although we perform computations on all packets in the network, we extract features by grouping them in flows. The FE computes median for a number of bytes and packets per flow in feature \# 9-12, 43-46, and 63-67. It computes the entropy, $H(F)$, for feature \# 8, 14, 16, 18, 20, 42, 48, 50, 54, 62, and 68 which is defined as follows:
\begin{align}
\label{eqn:ent}
H(F) = -\sum_{i=1}^{n}{\frac{f_i}{\sum_{j=1}^{n} f_j}\times{log_2{\frac{f_i}{\sum_{j=1}^{n} f_j}}}}
\end{align}
where set $F$=\{$f_1, f_2, ...,f_n$\} denotes the frequency of each distinct value. Once the FE extracts these features, it invokes the TC module implemented using SAE. It classifies traffic in one of the eight classes which includes one normal and seven types of DDoS attack classes based on TCP, UDP or ICMP vectors that adversaries launch either separately or in combinations.

\section{Experimental Set-up, Results, and Discussion}
\label{result}
\begin{table}
\centering
\begin{tabular}{ l|l|l|l }
  \hline
 \multicolumn{2}{c|}{\multirow{2}{*}{\textbf{Traffic class}}} & \multicolumn{2}{c}{\textbf{\# of records}}\\
 \cline{3-4}
 \multicolumn{2}{c|}{} & Training & Test \\
  \hline
  \hline
\multicolumn{2}{c|} {Normal (N)} & 49179 & 21076\\
\hline
\multirow{7}{*}{Attack} & TCP (T) & 5471 & 2344\\
 \cline{2-4}
{} & UDP (U)& 5273 & 2260\\
\cline{2-4}
{} &  ICMP (I)& 1602 & 686\\
 \cline{2-4}
{} &  TCP \& UDP (TU)  & 4694 & 2011\\
  \cline{2-4}
{} & TCP \& ICMP (TI) & 4739 & 2031\\
  \cline{2-4}
{} &  UDP \& ICMP (UI) & 4437 & 1902\\
  \cline{2-4}
{} &  All (A)& 5615 & 2407\\
  \hline
\end{tabular} 
\caption{Number of records in the training and test datasets for normal and different attack traffic}           
\label{table:dataset}
\end{table}  
To evaluate our system, we collected network traffic from a real network and a private network testbed. We discuss them along with the performance evaluation results.
\subsection{Experimental Set-up}
We used a home wireless network (HWN) connected to the Internet for normal traffic collection. The HWN comprised of around 12 network devices including laptops and smartphones. These devices were not uniformly active for all the time which led to variation in the traffic intensity. We saved HWN traffic of 72 hours in a Linux system using \texttt{tcpdump}~\cite{tcpdump} and port mirroring at a Wi-Fi access point. The traffic of first 48 hours were used as normal flows. The traffic of last 24 hours was mixed with the attack data that we collected separately and it was labeled as an attack in the presence of normal traffic. The collected traffic comprises data from web browsing, audio/video streaming, real-time messengers, and online gaming. To collect attack traffic, we created a private network in a segregated laboratory environment using VMWare ESXi host. The private network consists of 10 DDoS attacker and 5 victim hosts. We used \texttt{hping3}~\cite{hping} to launch different kinds of DDoS attacks with different packet frequencies and sizes. We launched one class of attack at a time so that it can be labeled easily while extracting features. After traffic collection in trace files, we created an SDN testbed on the same ESXi host similar to~\cite{hart} that consists of an SDN controller, an OF switch, and a network host using Ubuntu Linux systems. We used \texttt{POX}~\cite{pox}, a Python based controller, with our DDoS detection application running on it in the controller system and installed OpenvSwitch~\cite{ovs} in the switch to use it as an OF switch. In the host system, we used \texttt{tcpreplay}~\cite{replay} to replay traffic traces for normal and attack traffic one at a time. We saved features computed by the FE for each interval in dataset files for the training of TC module. We set the interval $60s$ in the $timer$ function to trigger the FE module for feature extraction. We divided the dataset files into training and test datasets. Table~\ref{table:dataset} shows the distribution of records in the dataset. Traffic features in the datasets are real-valued positive numbers. We normalize them using $max-min\ normalization$ shown in Eqn.~\ref{eqn:maxmin}, before passing them to the TC module.
\begin{align}
X_{norm} = \frac{x_i-x_{min}}{x_{max}-x_{min}},\ \forall x_i \in X
\label{eqn:maxmin}
\end{align}
\begin{align*}
x_{min} = \text{smallest value in}\ X \ \ \ \ \ \ \\
x_{max} = \text{largest value in}\ X \ \ \ \ \ \ \ \\
\end{align*}
\subsection{Results}
We evaluated the performance of our system on the datasets specified in Table~\ref{table:dataset} using parameters including \emph{accuracy, precision, recall, f-measure, ROC}. We use confusion matrix to calculate precision, recall, and f-measure. A confusion matrix, $M$, is an $N\times N$ matrix where $N$ is the number of classes. It represents the actual and predicted classes in such a way that columns are labeled for actual classes, and rows are labeled for predicted classes for all records. The diagonal elements of the matrix represent the true-positive (TP) for each class, sum of the matrix elements along a row excluding the diagonal element represents the number of false-positive (FP) for a class corresponding to that row, sum of the matrix elements along a column excluding the diagonal element represents the number of false-negative (FN) for a class corresponding to that column. Following are definitions of various performance parameters:
\begin{itemize}
\item \textit{Accuracy (A)}: percentage of accurately classified records in a dataset
\begin{align}
    A = \frac{Accurately\ classified\ records}{Total\ records}\times 100
\end{align}
\item \textit{Precision (P)}: number of accurately predicted records over all predicted records for a particular class. Using the confusion matrix, $M$, precision for each class, $j$, can be defined as follows:      
\begin{align*}
   {P_j} = \frac{TP_j}{TP_j\ +\ FP_j}\times 100\ \ \ \ \ \ \ \ \ 
\end{align*}  
\begin{align}
   {} = \frac{M_{j,j}}{M_{j,j}\ +\ \sum_{\substack{i=1\\i\ne j}}^N{M_{j,i}}} \times 100
 \end{align}  
\item \textit{Recall (R)}: number of accurately predicted records over all the records available for a particular class in the dataset. Using the confusion matrix, $M$, recall for each class, $j$, can be defined as follows:
\begin{align*}
{R_j} = \frac{TP_j}{TP_j\ +\ FN_j}\times 100\ \ \ \ \ \ \ \ \ \ 
\end{align*}  
\begin{align}
= \frac{M_{j,j}}{M_{j,j}\ +\ \sum_{\substack{i=1\\i\ne j}}^N{M_{i,j}}} \times 100
\end{align}
\item \textit{F-measure (F)}: It uses precision and recall for the holistic evaluation of a model and is represented as the harmonic mean of them. For each class, $j$, it is defined as follows:
\begin{align}
{F_j} = \frac{2\times P_j \times R_j}{P_j\ +\ R_j}\times 100
\end{align}
\item \textit{Receiver Operating Curve (ROC)}: It helps in visualizing a classifier's performance by plotting the true-positive rate against false-positive rate of the classifier. The area under the ROC gives an estimate of an average performance of the classifier. Higher the area, greater is the performance. 
\begin{figure}[hbtp]
\centering
\includegraphics[scale=0.55]{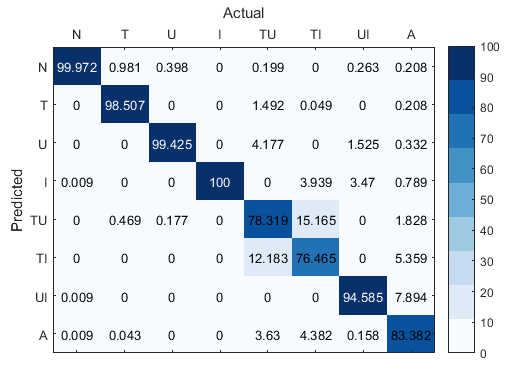}
\caption{Confusion matrix for 8-class classification in the SAE model}
\label{fig:8conf}
\end{figure}
\end{itemize}
\begin{figure}[hbtp]
\centering
\includegraphics[scale=0.22]{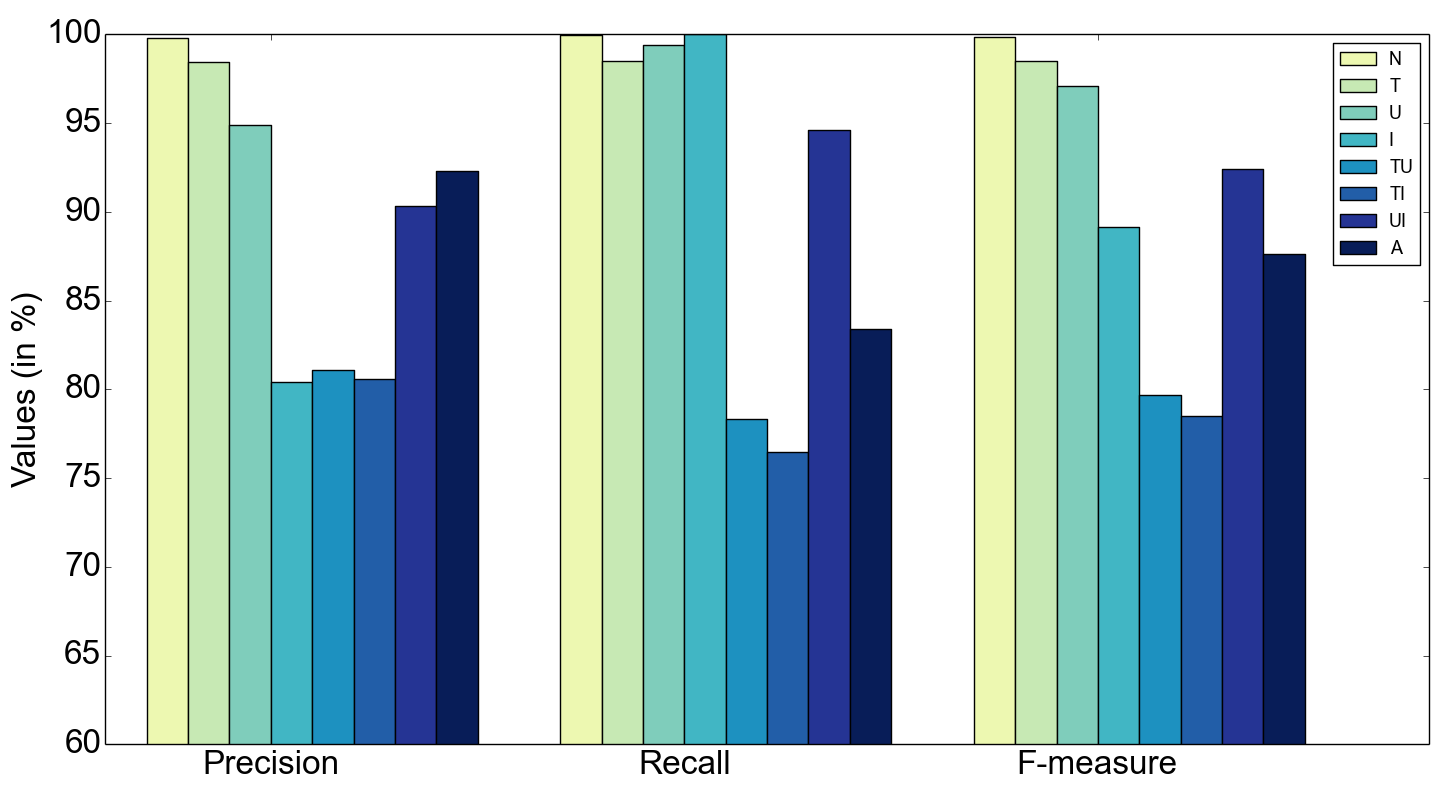}
\caption{Precision, recall, and f-measure for 8-class}
\label{fig:8prf}
\end{figure}
\begin{figure}[hbtp]
\centering
\includegraphics[scale=0.60]{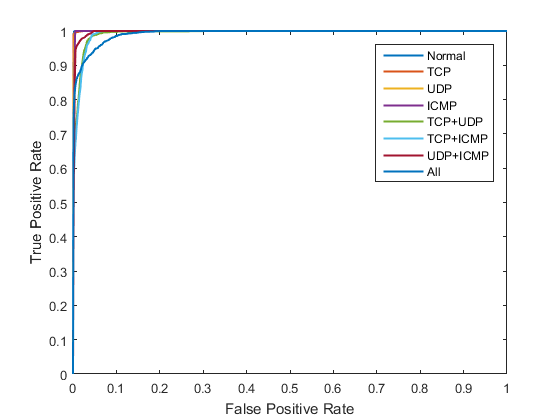}
\caption{ROC curve for 8-class classification}
\label{fig:8roc}
\end{figure}

We used the training dataset to develop the SAE classification model for the TC module and test dataset for performance evaluation. First, we developed the model for 8-class traffic classification including normal and seven kinds of DDoS attack that occur in combination with TCP, UDP, and ICMP based traffic. To make a better comparison, we also developed separate attack detection models with soft-max and neural network (NN) which are building blocks of SAE. As observed from Table~\ref{table:8comp}, the SAE model achieved better performance compared to the soft-max and neural network model in terms of accuracy. We computed precision, recall, f-measure for each traffic class. Figure~\ref{fig:8prf} shows their values which are derived from the confusion matrix shown in Figure~\ref{fig:8conf}. As seen from the figure, the model has f-measure value above 90\% for normal, TCP, UDP, and UDP with ICMP attacks traffic. It has comparatively low values of f-measure for TCP with ICMP and TCP with UDP attacks due to their classification of other kinds of attacks as observed from the Figure~\ref{fig:8conf}. However, it is observed from the same figure that the fraction of their classification as normal traffic is less than 0.2. Figure~\ref{fig:8roc} shows the ROC curve for 8 different classes. From the figure, we observe that the true positive rate is above 90\% with a false-positive rate of below 5\% for all kinds of traffic that results in the area under the ROC curve close to unity.
\begin{table}
\centering
\begin{tabular}{ l|l }
  \hline
  Method & Accuracy (in \%)\\
  \hline
  \hline
Soft-max & 94.30\\
\hline
Neural Network & 95.23\\
\hline
\textbf{SAE} & \textbf{95.65}\\
\hline
\end{tabular} 
\caption{Classification accuracy comparison among soft-max, neural network, and SAE based models}   \label{table:8comp}
\end{table}

\begin{figure}[hbtp]
\centering
\includegraphics[scale=0.60]{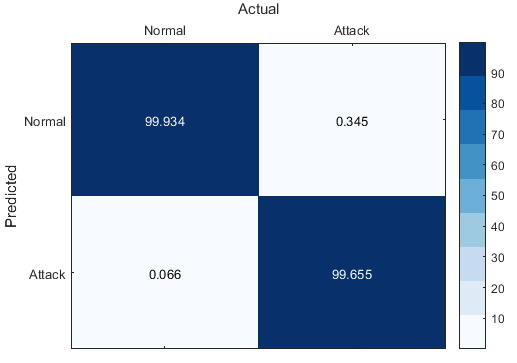}
\caption{Confusion matrix for 2-class classification}
\label{fig:2conf}
\end{figure}
\begin{figure}[hbtp]
\centering
\includegraphics[scale=0.225]{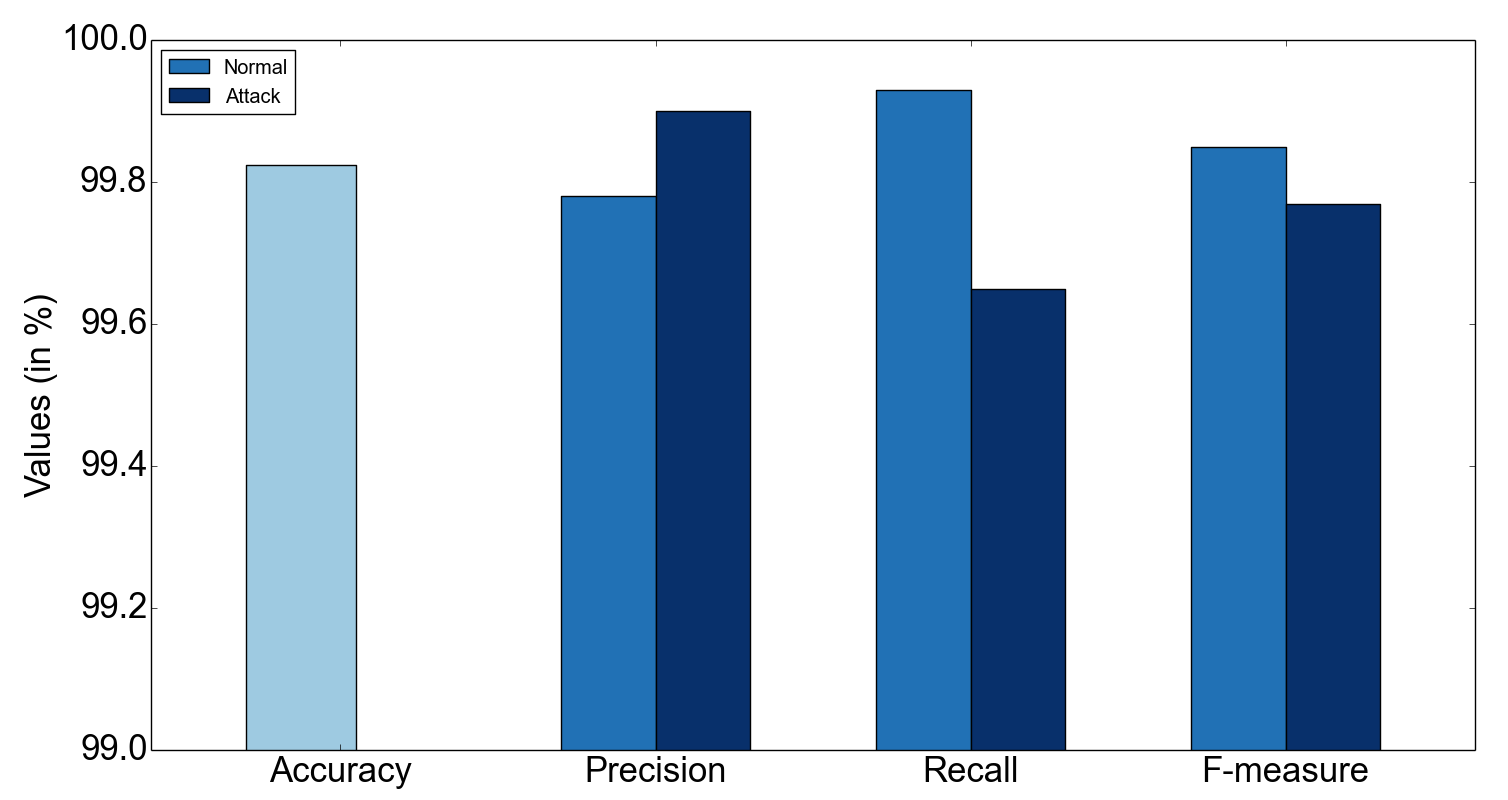}
\caption{Accuracy, precision, recall, and f-measure for 2-class classification}
\label{fig:2class}
\end{figure}
We evaluated our model for 2-class classification by considering all kinds of DDoS attacks as a single attack class to make a comparison with other works. Due to the unique nature of this work involving deep learning based attack detection in an SDN, and unavailability of existing literature in this specific domain, it was difficult to compare our work with other works. Figure~\ref{fig:2class} shows the performance for 2-class classification. The model achieved detection accuracy of 99.82\% with f-measure values as 99.85\% and 99.75\% for normal and attack classes, respectively, derived from the confusion matrix shown in Figure~\ref{fig:2conf}. On the contrary, the two closely related works~\cite{braga} and~\cite{mousavi}, achieved a detection accuracy of 99.11\% in data plane and 96\% in control plane, respectively. It should be noted that both of these works did not address attack detection in the other plane.

We also measured the computational time for training and classification in our model using a machine with Intel (R) Core i7 CPU @ 3.40 GHz processor and 16 GB RAM running Matlab 2016a on Windows 7. Table~\ref{tab:exectime} shows computational time for the training of 81,010 records and classification of 34,717 records specified in Table~\ref{table:dataset}.  
\begin{table}
    \centering
    \begin{tabular}{c|c}
    Training time &  Classification time \\
    \hline \hline
     524s & .0835s\\ 
    \end{tabular}
    \caption{Average computational time for the training and classification in the SAE model }
    \label{tab:exectime}
\end{table}

\subsection{Discussion}
With our DDoS detection system, we identify individual DDoS attack class and also determine whether an incoming traffic is normal or attack. A clear advantage in identifying each attack traffic type separately is enabling the mitigation technique to block only a specific type of traffic causing the attack, instead of all kinds of traffic coming towards the victim(s). Although we implemented a detection system, we separately extracted features for each host which has incoming traffic for an interval. Therefore, we can identify the hosts with normal traffic and the ones with attack traffic. Accordingly, the controller can install flow rules inside the switches to block the traffic for a particular host if it undergoes an attack. 

Our proposed system has a few limitations in terms of processing capabilities. The TCFI and FE modules collect every packet to extract features and are implemented on the controller for low false-positive in detection. However, this approach may limit the controller's performance in large networks. We can overcome it by adopting a hybrid approach that can either use flow sampling or individual packet capturing based on the observed traffic in the organizational network. Another approach that could be employed to handle DDoS attacks in the data plane is to deploy the TCFI and FE modules in another host, send all packets to it instead of the controller for features processing, and then periodically notify the controller with extracted features for the TC module. To reduce the time in feature extraction, we can also apply distributed processing similar to our another previous work~\cite{maroof}.
\section{Conclusion}
\label{conclusion}
In this work, we implemented a deep learning based DDoS detection system for multi-vector attack detection in an SDN environment. The proposed system identifies individual DDoS attack class with an accuracy of 95.65\%. It classifies the traffic in normal and attack classes with an accuracy of 99.82\% with very low false-positive compared to other works. In the future, we aim to reduce the controller's bottleneck and implement an NIDS that can detect different kinds of network attacks in addition to DDoS attack. We also plan to use deep learning for feature extraction from raw bytes of packet headers instead of feature reduction from the derived features in future NIDS implementation.
\bibliographystyle{spmpsci}
\bibliography{arxiv}
\end{document}